\documentclass[9pt,twocolumn,twoside]{opticajnl}
\journal{opticajournal} 
\usepackage{multirow, makecell}
\setboolean{shortarticle}{true}

\usepackage{lineno}

\title{Rapid adiabatic couplers with arbitrary split ratios for broadband DWDM interleaver application}

\author[1, $\dagger$]{Daehan Choi}
\author[1, $\dagger$]{Woo-Joo Kim}
\author[1, *]{Young-Ik Sohn}

\affil[1]{School of Electrical Engineering, Korea Advanced Institute of Science and Technology (KAIST), Daejeon 34141, Republic of Korea}
\affil[$\dagger$]{These authors contributed equally to this work.}
\affil[*]{youngik.sohn@kaist.ac.kr}

\begin{abstract}
We experimentally demonstrate a compact and broadband rapid adiabatic couplers (RACs) with arbitrary power split ratios, achieved through the combination of translational offset and waveguide width control. Fabricated RACs of four different target split ratios show power splitting within ±3\% of the design target over a 160 nm wavelength range. Using these RACs, we implement an 8-channel dense wavelength division multiplexing (DWDM) interleaver exhibiting < –20 dB crosstalk for the center 8 channels with flat-top passbands. Over a broader wavelength range, the design maintains crosstalk below –10 dB across more than 40 channels with 100 GHz spacing, demonstrating the broadband capability and scalability of RAC-based photonic integrated circuits.
\end{abstract}

\setboolean{displaycopyright}{false} 

\begin{document}

\maketitle
Photonic integrated circuits (PICs) are rapidly evolving to support systems of increasing complexity and scale, driven by demands in high-speed data communications, signal processing, and emerging quantum technologies \cite{Bogaerts2020, shekhar2024}. As these systems grow, efficient routing and power splitting become more essential, making passive photonic components critical to scalable integration. Among these, couplers play a central role in enabling reliable signal distribution. However, conventional interference-based implementations, such as directional couplers and multimode interferometers, suffer from limited spectral bandwidth and high sensitivity to fabrication variations. 

To address these challenges, adiabatic couplers have been adopted for their broadband performance and robustness to fabrication variations \cite{Ramadan1998}. By gradually varying the waveguide geometry along the propagation axis, they allow smooth mode evolution without significant intermodal coupling. This design strategy leads to robust performance across a wide range of wavelengths, making them particularly suitable for wavelength division multiplexing (WDM) applications. Nonetheless, their relatively long device length---often several hundred \textmu m---remains a limitation for compact integration.

Recently, an improved design method called rapid adiabatic coupler (RAC) was introduced, in which an additional geometric degree of freedom along the coupler can effectively suppress the excitation of undesired higher-order eigenmodes \cite{Cabanillas2021, Cabanillasthesis}. Hence, RAC designs minimize unwanted coupling and enable adiabatic mode evolution in substantially shorter devices. Meanwhile, studies on conventional adiabatic couplers have demonstrated that adjusting the final waveguide width enables arbitrary split ratios \cite{Mao2019}. Inspired by these two results, we further extend initial demonstrations of RACs, which were applied only to a 50:50 power-splitting ratio, to devices that can achieve arbitrary split ratios by applying novel design methodologies. This improvement from the previous RAC design can significantly enhance their applicability in diverse photonic systems.

We demonstrate the practical versatility of our RAC design by implementing chip-scale dense wavelength division multiplexing (DWDM) interleavers. One of the known approaches to PIC-based DWDM is to use lattice filters, where directional couplers of different split ratios are typically used to define their spectral responses \cite{Horst2013}. However, these directional couplers exhibit inherent wavelength sensitivity, limiting bandwidth performance. Replacement with broadband RACs alleviates this constraint, providing uniform coupling performance across an extensive wavelength range. By significantly shortening coupler lengths and precisely controlling split ratios, our design enables broadband and highly scalable lattice-filter-based DWDM interleavers, significantly enhancing integration density and robustness in photonic circuits. 

\begin{figure}[ht]
\centering
\fbox{\includegraphics[width=0.95\linewidth, keepaspectratio]{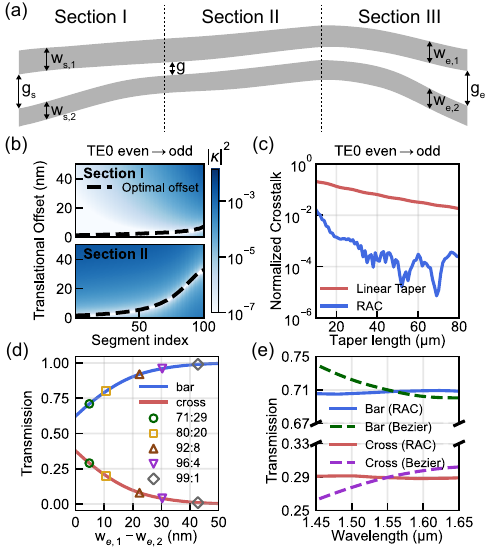}}
\caption{
(a) Schematic of RAC structure. 
(b) Power coupling ($|\kappa|^2$) between the first two TE modes versus translational offset in Sections I and II; black dashed line shows the selected optimal offset for each segment. 
(c) Simulated TE\textsubscript{0} even-to-odd crosstalk in Section II as a function of length. The red line represents a conventional adiabatic coupler with linear tapers, while the blue line represents the proposed RAC. 
(d) Output transmission as a function of width difference $\text{w}_\text{e,1} - \text{w}_\text{e,2}$; markers indicate target split ratios. 
(e) Simulated spectra of 71:29 output bends designed using translational offset optimization (solid) and conventional Bezier S-bends (dashed).
}
\label{fig:1}
\end{figure}

The RACs are implemented on a silicon photonic platform with a 220-nm-thick device layer. The corresponding layout is shown in Fig. \ref{fig:1}(a), where the coupler is divided into three sections. In Section I, the structure begins at the initial gap ($\text{g}_\text{s}$), where the coupling between two waveguides is negligible. This gap is gradually narrowed to $\text{g}$ for Section II, while the waveguide widths ($\text{w}_\text{s,1}$ and $\text{w}_\text{s,2}$) remain unchanged. Section II is where mode evolution takes place, with waveguide widths tapering from $\text{w}_\text{s,1}$ and $\text{w}_\text{s,2}$ to $\text{w}_\text{e,1}$ and $\text{w}_\text{e,2}$, respectively, while the gap ($\text{g}$) is maintained. By the end of Section II, we obtain the desired power split ratio. Finally, in Section III, the gap is again increased to a value ($\text{g}_\text{e}$) while the power split ratio of the two waveguides is maintained.

The initial waveguide widths, $\text{w}_\text{s,1}$ and $\text{w}_\text{s,2}$, were defined relative to a 500 nm bus waveguide ($\text{w}_\text{bus}$), with $\Delta\text{w}_\text{s}$ denoting their difference: $\text{w}_\text{s,1} = \text{w}_\text{bus} + \Delta\text{w}_\text{s}/2$ and $\text{w}_\text{s,2} = \text{w}_\text{bus} - \Delta\text{w}_\text{s}/2$. If $\Delta\text{w}_\text{s}$ is too large, the narrower waveguide suffers from excessive optical loss. If it is too small, the unwanted coupling at the end of Section I increases, requiring a longer adiabatic transition. After several iterations, we chose to use $\Delta\text{w}_\text{s}$ = 160 nm, which is nearly optimal.

To determine the appropriate value for $\text{g}_{\text{s}}$, we used a finite difference eigenmode (FDE) solver to analyze the coupling between two waveguides. With waveguide widths of 580 and 420 nm, we found that $\text{g}_{\text{s}} = 700$ nm results in negligible coupling. In contrast, the gap $\text{g}$ used in Section II should be small enough to support effective power transition and allow for shorter device length. Although smaller gaps enhance coupling, they introduce fabrication challenges, especially cladding voids that can degrade device performance \cite{shiran2020, warshavsky2025}. Considering both optical performance and fabrication reliability, we chose $\text{g} = 100$ nm.

Following the approach in \cite{Cabanillas2021, Cabanillasthesis}, we added a geometric degree of freedom to control mode evolution along the coupler by applying curvature through translational offsets perpendicular to the propagation direction in Sections I and II. This approach effectively suppresses coupling between the first and second TE eigenmodes, which are even-like and odd-like supermodes, enabling shorter coupler designs. For the 50:50 baseline design, we set equal output widths at $\text{w}_\text{e,1}=\text{w}_\text{e,2}=500\ \text{nm}$. Sections I and II were each discretized into 100 segments, and the optimal offset for every segment was calculated with the FDE solver. Fig. \ref{fig:1}(b) shows the relationship between translational offset and coupling between the first and second TE modes. Afterwards, the final coupler geometry was obtained by smoothly interpolating these discrete offsets. For further optimization, the segment lengths were adaptively stretched in regions with stronger coupling and compressed in weaker regions. As shown in Fig. \ref{fig:1}(c), this design achieves significantly shorter coupler lengths compared to a conventional linear taper-based approach while maintaining low intermodal crosstalk. 

Although the baseline design achieves a 50:50 power split, arbitrary split ratios require introducing asymmetry along the transition from Section II to Section III. Following the method in \cite{Mao2019}, we varied the widths of the output waveguides to induce an effective index difference, which shifts the relative fraction of modal power in each waveguide: the even-like fundamental mode becomes more confined in the wider waveguide, while the odd-like second mode becomes more confined in the narrower one. Section III must then widen the gap while maintaining this power distribution and avoiding crosstalk. 

To determine the widths of two waveguides at the end of Section II to achieve target split ratios, we performed a parameter sweep around the 500 nm bus width---setting $\text{w}_\text{e,1} = 500\ \text{nm} + \Delta\text{w}_\text{e}/2$ and $\text{w}_\text{e,2} = 500\ \text{nm} - \Delta\text{w}_\text{e}/2$---and obtained optimal $\Delta\text{w}_\text{e}$. The optimized values of $\Delta \text{w}_\text{e}$ for each target split ratio are listed in Table S1 of the Supplement. Fig. \ref{fig:1}(d) shows the corresponding power-split curves as a function of $\Delta\text{w}_\text{e}$, with the target ratios indicated by markers. 

After obtaining two waveguide widths, $\text{w}_\text{e,1}$ and $\text{w}_\text{e,2}$, we design Section III to only perform the fan-out function without disturbing the established power distribution of the two waveguides at the end of Section II. In the case of a 50:50 coupler, Section III typically consists of two symmetric Bezier S-bends that preserve mirror symmetry between the waveguides. As a result, the even and odd modes remain eigenmodes throughout Section III, and there is no coupling between them. 

However, when asymmetric waveguide widths are introduced to achieve imbalanced power splitting, the same Bezier S-bends with different waveguide widths break mirror symmetry and increase modal overlap between the two supermodes. This overlap induces residual coupling in Section III that alters the split ratio given at the end of Section II. To mitigate this issue, we applied the same translational offset technique used in Sections I and II to Section III. In this way, Section III can only spatially separate two waveguides while keeping the power split ratios achieved at the end of Section II. Fig. \ref{fig:1}(e) compares the spectral characteristics of 71:29 output bends for the two cases: Section III waveguides were implemented with conventional Bezier S-bends versus those designed using offset optimization. The result shows that the latter provides wider bandwidth performance by better suppressing residual coupling. 

\begin{figure}[ht]
\centering
\fbox{\includegraphics[width=0.95\linewidth, keepaspectratio]{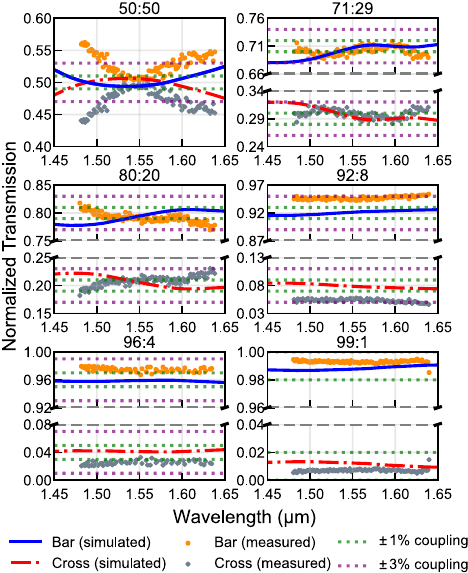}}
\caption{Simulated and experimental results of each RAC.}
\label{fig:2}
\end{figure}

The solid blue line and the dashed-dot red lines in Fig. \ref{fig:2} show the simulated spectral response of the entire RAC structure for each target split ratio, which confirms the broadband behavior. The simulations were performed using the 3D finite-difference time-domain (FDTD) method. The corresponding total lengths of the RACs are summarized in Supplementary Table S2.

To verify the power splitting ratios experimentally, asymmetric Mach-Zehnder interferometers (AMZI) were constructed using two couplers of the same type. The splitting ratio of each coupler was extracted from the extinction ratio of the AMZI, assuming that the two couplers are identical \cite{Bogaertsreview}. TE-pass filters were included at the input of each AMZI to only excite the fundamental TE mode, designed based on the method described in \cite{zafar2018} to suppress higher-order and TM-like modes.

The scatter plot in Fig. \ref{fig:2} presents the extracted power splitting ratios derived from the measured spectral responses of the AMZIs. At the center wavelength of 1550 nm, all six designs exhibit split ratios within $\pm3$\% of their respective targets, showing good agreement with the design values. Furthermore, four of the six designs, other than 50:50 and 92:8 couplers, maintain this $\pm3$\% error throughout the entire 160 nm wavelength range tested (1480–1640 nm), which was only limited by the CW laser tuning range. The remaining two designs, 50:50 and 92:8, satisfy the $\pm3$\% condition in narrower ranges: 1512.8-1588.6 nm and 1480-1599 nm, respectively. These results confirm the effectiveness of the proposed design method in realizing compact, broadband adiabatic couplers with accurate power splitting. With more advanced fabrication processes offering tighter control over waveguide width, even higher precision can be expected. 

\begin{table}[ht]
\caption{Summary of recent experimental results for imbalanced adiabatic couplers.}
  \label{tab:imb-ac-results}
  \centering
\begin{tabular}{cccc}
\hline
\makecell{Power\\split ratio} & \makecell{Mode evolution\\region length} & \makecell{$\pm3\%$\\Bandwidth} & Ref. \\
\hline
78:22 & 200 \textmu m & 100 nm & \cite{Mao2019}\\
78:22 & 42 \textmu m & 70 nm & \cite{Chen2024}\\
71:29 & 70.814 \textmu m & > 160 nm & This work\\
80:20 & 42.85 \textmu m & > 160 nm & This work\\
92:8 & 29.04 \textmu m & > 119 nm & This work\\
96:4 & 22.95 \textmu m & > 160 nm & This work\\
99:1 & 18.97 \textmu m & > 160 nm & This work\\
\hline
\end{tabular}
\end{table}

To our knowledge, our 80:20 RAC achieves a length comparable to or shorter than previously reported 78:22 adiabatic couplers \cite{Mao2019, Chen2024}, while significantly improving spectral performance. For reference, Table \ref{tab:imb-ac-results} summarizes recent experimental results on unbalanced adiabatic couplers, comparing bandwidth and coupler length. To ensure a fair comparison among known results, we have used only the length of the mode evolution region, which corresponds to Section II of Fig. \ref{fig:1}(a). We limit the comparison to this region because $\text{g}_\text{s}$ and $\text{g}_\text{e}$ in Sections I and III can be chosen case by case---e.g., to match input bus waveguides or meet packaging clearances---so their lengths can vary widely across reports and are not directly comparable.

\begin{figure}[t]
\centering
\fbox{\includegraphics[width=0.96\linewidth, keepaspectratio]{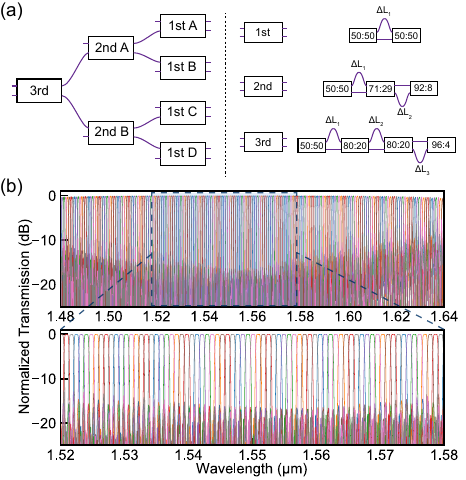}}
\caption{
(a) Schematic of DWDM interleaver and first-, second- and third-order AMZIs. 
(b) Simulated spectrum of DWDM interleaver. Simulation was performed from 1480-1640 nm. The magnified figure for the range 1520-1580 nm was separately depicted for direct comparison with the experimental data.}
\label{fig:3}
\end{figure}

Based on well-known lattice filter theories \cite{Horst2013}, we designed third-, second-, and first-order AMZIs by connecting 50:50, 71:29, 80:20, 92:8, and 96:4 RACs previously described. Fig. \ref{fig:3}(a) shows the configuration of the DWDM interleaver used in our study. To achieve 100 GHz channel spacing, a cascading structure of AMZIs with 200 GHz, 400 GHz, and 800 GHz free spectral range (FSR) is employed. Phase delay arm lengths were adjusted to align the center wavelengths of each channel with the ITU grid. The optimized values used in each AMZI stage are summarized in Table S3 of the Supplement. Fig. \ref{fig:3}(b) shows the simulated spectral response of the designed DWDM interleaver.

\begin{figure}[htbp]
\centering
\fbox{\includegraphics[width=0.96\linewidth, keepaspectratio]{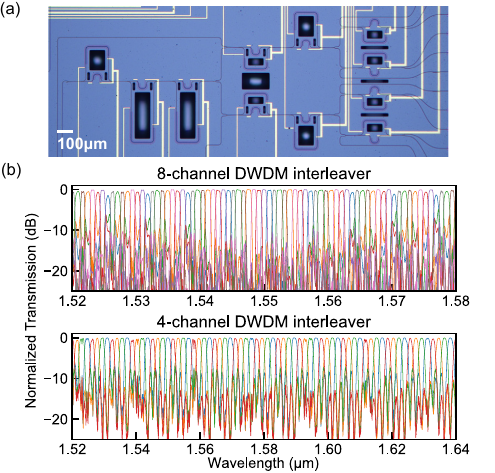}}
\caption{
(a) Microscope image of DWDM interleaver. 
(b) Experimental spectra of the 8-channel DWDM interleaver with 100 GHz channel spacing (top) and a 4-channel interleaver with 200 GHz spacing (bottom) implemented using part of the same device. Each color represents a different output channel.
}
\label{fig:4}
\end{figure}

Fig. \ref{fig:4}(a) shows a microscope image of the fabricated device. The devices were fabricated by Applied Nanotools Inc. using electron beam lithography. TiW heaters were included on each MZI arm to compensate for phase errors caused by fabrication. Similarly to the AMZI devices described in the previous section, a TE-pass filter was added to the input port of the interleaver. 

The top plot in Fig. \ref{fig:4}(b) presents the measured transmission spectra of the 8-channel DWDM interleaver. The measured crosstalk is less than –20 dB for more than eight channels around 1550 nm, and the –0.5 dB flat-top bandwidth of each channel exceeds 30 GHz (0.24 nm). Due to the broadband nature of our coupler, the interleaver exhibits low crosstalk (< –10 dB) across more than 40 channels within the 1530-1570 nm wavelength range. Current performance is primarily limited by one block of the entire circuit, second-order AMZI A, where phase mismatch arises primarily from fabrication-induced width errors in the long differential arm. The accumulated phase error could not be fully corrected by the integrated heater, because the available phase tuning range did not reach the full $2\pi$. This limitation, in principle, can be mitigated by adopting a dispersion-aware AMZI arm design, as demonstrated even under fabrication imperfections that caused group index variations \cite{xie2020}.

To further evaluate the broadband potential of the architecture, a partial configuration that excludes AMZI A was experimentally tested. In the configuration shown in Fig. \ref{fig:3}(a), the unconnected input port of the second-order AMZI B is utilized, while the two connected first-order AMZIs remain unchanged. This partial structure, consisting of one second-order AMZI B and two first-order AMZI C and D, functions as a second-order lattice filter with 200 GHz channel spacing. In this configuration, crosstalk levels below –10 dB were observed around the 1525–1640 nm wavelength range, covering over 60 channels with 200 GHz spacing except for one channel that slightly exceeded -10 dB near 1558 nm. These results highlight the broadband performance of the proposed RAC-based interleaver design and its scalability, demonstrating its suitability for compact and flexible PIC implementations.

In summary, we demonstrated RACs with arbitrary split ratios using a combination of translational offset engineering and waveguide width control. The designed couplers feature compact lengths and broadband characteristics, with experimentally verified split ratios within ±3\% at the target wavelength of 1550 nm. Most of the designs with different split ratios show a similar level of accuracy across the 160-nm wavelength span, confirming their wideband performance of the proposed devices. Based on the reliability of these individual RACs with different split ratios, we implemented an 8-channel DWDM interleaver that showed low crosstalk (< –20 dB) in the central region and maintained < –10 dB crosstalk over 40 channels with 100 GHz spacing. These results highlight the potential of RACs as versatile building blocks for scalable and dense photonic integrated circuits. Furthermore, their broadband behavior and the feasibility of an arbitrary split ratio suggest that the proposed RACs are also well suited for broadband CWDM systems, where wider channel spacing benefits from the use of power splitters that are both spectrally flat and robust against fabrication variation.

Although we have chosen the DWDM interleaver as a representative example for this work, RACs with arbitrary split ratios have great potential to become crucial building blocks of various PIC circuits, particularly those with a large number of elements. The fact that RACs are robust against fabrication errors is the key reason for their importance in large-scale circuits. For example, as the circuit becomes larger, it is very likely that more power splitters will be used. To ensure that the circuit works as designed, it is essential that each component is robust against fabrication errors, and that is where RACs or similar robust devices can be employed \cite{Hamerly2022}.

\begin{backmatter}
\bmsection{Funding} National Research Foundation of Korea (RS-2021-NR061364, RS-2022-NR068818, RS-2024-00442762, NRF-2022M3E4A1077013, NRF-2022M3H3A1085772); Institute of Informations \& Communications Technology Planning \& Evaluation (RS-2024-00397959) 


\bmsection{Disclosures} The authors declare no conflicts of interest.

\bmsection{Data availability} Data underlying the results presented in this paper are not publicly available at this time but may be obtained from the authors upon reasonable request.

\bmsection{Supplemental document} See Supplement 1 for additional design parameters.

\end{backmatter}

\bibliography{references}


\end{document}